\documentclass[a4paper,11pt]{article}
\pdfoutput=1 

\usepackage{jheppub} 

\usepackage[T1]{fontenc} 

\title{\boldmath The effect of quantum correction on Hawking radiation for Schwarzschild black holes}

\author{Yang Liu}

\affiliation[a]{School of Physics and Astronomy, University of Nottingham, Nottingham NG7 2RD, UK}
\affiliation[b]{Nottingham Centre of Gravity, University of Nottingham, Nottingham NG7 2RD, UK}

\emailAdd{yang.liu@nottingham.ac.uk}

\abstract{We investigate the effect of quantum correction on Hawking radiation for Schwarzschild black holes. We consider Hawking temperature and entropy to order $G^2$ and find that the area law of black holes should be modified. We think of Hawking radiation as tunneling and find that at certain frequency $\omega$ the radiation spectrum of black holes can be pure blackbody spectrum. Therefore, it is possible to break down information conservation. In order to ensure information conservation at any energy scales, we suggest that black holes cannot release all information they have. We briefly discuss the bound on greybody factors for Schwarzschild black holes as well. Since the modification of horizon is very tiny, the bound on greybody factors has a very small difference between classical metric and quantum corrected metric.}

\begin{document} 
\maketitle
\flushbottom

\section{Introduction}
In 1974, since Hawking's original discovery [1,2] has showed that black holes can spontaneously emit particles at a temperature inversely proportional to their mass. Black holes, as a fascinating and elegant object, have been increasingly popular in classical and quantum gravity theories [3]. Meanwhile, a plenty of works [4–8] devoted to calculating the Hawking temperature and radiation in order to explore a possible quantum gravity theory [3]. For example, in refs.[4] and [5], Page has calculated the particle emission rates from a nonrotating and rotating black hole, respectively. In 1999, Wilczek and Parikh [6] showed that Hawking radiation can be viewed as a tunnelling process, in which particles pass through the contracting horizon of the black hole, lending formal justification to the most intuitive picture we have of black hole evaporation [9].\\
Other studies have generalized these results to other cases, such as charged black holes [3,10], Einstein-Gauss-Bonnet de Sitter black hole [11], etc. Here we only list some recent studies. Avijit Chowdhury and Narayan Banerjee [10] considered Hawking radiation from a charged spherical black hole with scalar hair and showed that the scalar and electric charges contribute oppositely to the greybody factor and the sparsity of the Hawking radiation cascade. Yan-Gang Miao and Zhen-Ming Xu [3] investigated the Hawking radiation cascade from the five-dimensional charged black hole with a scalar field coupled to higher-order Euler densities in a conformally invariant manner. The authors found that the Hawking radiation cascade from this five-dimensional black hole is extremely sparse and the charge enhances the sparsity of the Hawking radiation, while the conformally coupled scalar field reduces this sparsity. Cheng-Yong Zhang, Peng-Cheng Li and Bin Chen [11] studied the greybody factors of the scalar fields in spherically symmetric Einstein-Gauss-Bonnet–de Sitter black holes in higher dimensions. They derived the greybody factors analytically for both minimally and nonminimally coupled scalar fields and found that the nonminimal coupling may suppress the greybody factor and the Gauss-Bonnet coupling could enhance it, but they both suppress the energy emission rate of Hawking radiation.\\
All articles, however, assume the metrics of black holes are classical, i.e., the quantum effect on metrics have been neglected. Due to effective field theory (EFT) methods, the quantum corrected metric of Schwarzschild black holes has been obtained recently [12]. Since the seminal work of Weinberg in 1979 [13], much progress has been made in quantum gravity using effective field theory methods [14–20]. Although finding a consistent quantum gravity at all energy scales is still an extremely difficult problem at the moment, EFT methods can be applied at energies below the Planck mass and enables calculations in quantum gravity which are model independent [12,21]. The only two requirements are that Lorentz invariance is a fundamental symmetry and general coordinate invariance is the correct symmetry at low energy scale [12]. The quantum gravitational effective action contains two parts: local and nonlocal operators. The Wilson coefficients of the local operators cannot be calculated if we have no knowledge of the ultraviolet complete theory of quantum gravity, however, the Wilson coefficients of the non-local operator can be obtained from first principles and only depend on the infrared physics which we have understood very well [21]. More related details can be found in refs.[12,21].\\ 
The main aim of the original paper [12] is to describe the metric inside and outside a star, taking into account quantum effects. However, the authors also proposed that the exterior metric may provide a good model of quantum black holes [12]. Therefore, one direct following research of ref.[12] is to consider the quantum corrections to Hawking temperature and radiation power, which is the research topic of this article. In section 2, we briefly review the quantum corrected metric. In section 3, the quantum correction of Hawking temperature and entropy for Schwarzschild black holes to order $G^2$ have been obtained. In section 4, we think of Hawking radiation as tunneling. In section 5, we determine the bound on greybody factors for Schwarzschild black holes to the lowest order. In section 6, the results of this article have been discussed. We take $c=k_B=\hbar=1$ throughout. $G$ is the Newton gravitational constant.

\section{Brief review of the quantum corrected metric}
In this section, we will review the quantum corrected metric to order $G^2$ obtained from Effective Field Theory (EFT). The quantum effective action is given by [12]:
\begin{equation}\label{eq:2.1}
\Gamma [g] = \Gamma_L [g] + \Gamma_{NL} [g],  
\end{equation}
where the local part of the action is given by [12]
\begin{equation}\label{eq:2.2}
\Gamma_L [g] = \int d^4x \sqrt{g} [\frac{R}{16 \pi G} + c_1(\mu)R^2 + c_2(\mu) R_{\mu\nu} R^{\mu\nu} + c_3(\mu) R_{\mu\nu\alpha\beta} R^{\mu\nu\alpha\beta}]  
\end{equation}
and the non-local part of the action by [12]
\begin{equation}\label{eq:2.3}
\Gamma_{NL} [g] = - \int d^4 x \sqrt{g} [\alpha R \ln(\frac{\Box}{\mu^2})R + \beta R_{\mu\nu} \ln(\frac{\Box}{\mu^2})R^{\mu\nu} + \gamma R_{\mu\nu\alpha\beta} \ln(\frac{\Box}{\mu^2})R^{\mu\nu\alpha\beta}]
\end{equation}
Here, the parameters $\alpha$, $\beta$ and $\gamma$ are non-local Wilson coefficients for fields. The values of the parameters for different fields have been listed in ref.[12]. \\
In ref.[12], the quantum corrected metric for a static and spherical star and black hole has been obtained, which can be written as:
\begin{equation}\label{eq:2.4}
ds^2 = -f(r) dt^2 + g(r) dr^2 + r^2 d\Omega^2
\end{equation}
where $d\Omega^2 = d\theta^2 +(\sin \theta)^2 d\phi^2$. Outside the radius of star $R_s$ (i.e., $r > R_s$), the metric functions are given by
\begin{equation}\label{eq:2.5}
f(r) = 1 - \frac{2GM}{r} + \alpha_e(r)
\end{equation}
\begin{equation}\label{eq:2.6}
g(r) = \left(1 - \frac{2GM}{r}\right)^{-1} + \beta_e(r)
\end{equation}
where
\begin{equation}\label{eq:2.7}
\alpha_e(r) = \tilde{\alpha} \frac{2G^2M}{R^3_s} [2\frac{R_s}{r} + \ln (\frac{r-R_s}{r+R_s})] + O(G^3)
\end{equation}
\begin{equation}\label{eq:2.8}
\beta_e(r) = \tilde{\beta} \frac{2G^2M}{r (r^2 - R^2_s)} + O(G^3)
\end{equation}
with
\begin{equation}\label{eq:2.9}
\tilde{\alpha} = 96\pi(\alpha + \beta +3\gamma)
\end{equation}
\begin{equation}\label{eq:2.10}
\tilde{\beta} = 192 \pi (\gamma - \alpha)
\end{equation}
and $M$ is the mass of the black hole. We should emphasize that the corrections to the metric should include two parts: one is from the local part of $(2.1)$ (i.e., eq.$(2.2)$), while the other part is from the non-local part of $(2.1)$ (i.e., eq.$(2.3)$) [12]. According to ref.[12], the local part has no contribute. In particular, the corrections outside the star are trivially $0$ while the corrections inside the star turn out to be order $O(G^3)$, and thus subleading [12]. The two terms $\tilde{\alpha}$ and $\tilde{\beta}$ are the contributions of non-local part of $(2.1)$, i.e., eq.$(2.3)$. Therefore, we cannot expand the first term of $(2.6)$ in powers of $G$ to obtain the term $\tilde{\beta}$. Moreover, we should point out that the true perturbation parameters are the inverse of the radius of the curvature in units of Planck length and the compactness of the star, which are dimensionless [21].\\
By introducing the tortoise coordinate [21]
\begin{equation}\label{eq:2.11}
r_{\ast} = \int^{r} \sqrt{\frac{g(r')}{f(r')}} dr' 
\end{equation}
The metric can be rewritten as [21]
\begin{equation}\label{eq:2.12}
ds^2 = -f(r) (dt^2 + dr_{\ast}^2) + r^2 d\Omega^2 
\end{equation}
In the exterior region, the tortoise coordinate is [21]
\begin{equation}\label{eq:2.13}
r_{\ast}(r) = r + 2GM \ln (\frac{r}{2GM} - 1) + \frac{1}{2} \int_{r}^{\infty} [\alpha_e (r') - \beta_e (r')] dr' + C, 
\end{equation}
where $C$ is an integration constant.

\section{Quantum correction of Hawking temperature and entropy for Schwarzschild black holes}
In this section, we study the semiclassical tunnelling of particles through the horizon of Schwarzschild black holes, in line with the analysis of ref.[9]. Further discussion can be seen in refs.[22-25]. On the one hand, the emitted rate of particles $\Gamma$ is given by
\begin{equation}\label{eq:3.1}
\Gamma \sim \exp \left(-2 Im S \right) 
\end{equation}
where $S$ is the tunnelling action of particles and $Im S$ is the imaginary part of the tunnelling action $S$. On the other hand, based on the Planck radiation law, the emitted rate $\Gamma$ of particles with frequency $\omega$ can be written as
\begin{equation}\label{eq:3.2}
\Gamma \sim \exp \left(-\omega/T_{BH} \right)
\end{equation}
Therefore, the temperature at which the black hole radiates can be read off:
\begin{equation}\label{eq:3.3}
T_{BH} = \frac{\omega}{2 Im S} 
\end{equation}
In section 3, we will study the process in Painlevé-Gullstrand coordinates. Firstly, we need to rewrite the corrected metric in Painlevé-Gullstrand coordinates to order $G^2$. We define
\begin{equation}\label{eq:3.4}
t_r =t - a(r)
\end{equation}
where $a(r)$ is a function of radial coordinate $r$. Then we have
\begin{equation}\label{eq:3.5}
dt^2 = d t^2_r + a'^{2}(r) dr^2 + 2 a' dt_r dr
\end{equation} 
where $a' = da(r)/dr$. The time coordinate $t_r$, just as for Schwarzschild time $t$, corresponds to the time measured by a stationary observer at infinity [9]. Inserting eq.$(3.5)$ into eq.$(2.4)$, then the metric can be rewritten as 
\begin{equation}\label{eq:3.6}
ds^2 = -f(r) d t^2_r - 2f(r)a' dt_r dr + \left(g(r) - f(r) a'^2 \right) dr^2 + r^2 d \Omega^2
\end{equation}
We require
\begin{equation}\label{eq:3.7}
g(r) - f(r) a'^{2} =1
\end{equation}
then we have $a'^{2} = \frac{g(r) -1}{f(r)}$. We choose $a' = - \sqrt{\frac{g(r) -1}{f(r)}}$. Then
\begin{equation}\label{eq:3.8}
-2 f(r) a' = 2 \sqrt{f(r) \left(g(r) -1\right)}
\end{equation} 
For classical Schwarzschild metric, $f(r) = 1- \frac{2GM}{r}$ and $g(r) = \left(1 - \frac{2GM}{r}\right)^{-1}$, then we have
\begin{equation}\label{eq:3.9}
ds^2 = - \left(1- \frac{2GM}{r} \right) dt^2_r + 2 \sqrt{\frac{2GM}{r}} dt_r dr + dr^2 + r^2 d\Omega^2
\end{equation} 
For the quantum corrected Schwarzschild metric, $f(r)$ and $g(r)$ have been given in eqs.$(2.5)$ and $(2.6)$, then we have
\begin{equation}\label{eq:3.10}
f(r) \left(g(r) -1\right) = \frac{2GM}{r} + \frac{2 \tilde{\beta} G^2 M}{r(r^2 - R^2_s)} + O(G^3)
\end{equation}
and
\begin{equation}\label{eq:3.11}
-2 f(r) a' = 2 \sqrt{f(r) \left(g(r) -1\right)} = 2 \sqrt{\frac{2GM}{r} + \frac{2 \tilde{\beta} G^2 M}{r(r^2 - R^2_s)}}
\end{equation}
Therefore, to order $G^2$, the quantum corrected Schwarzschild metric can be written as
\begin{equation}\label{eq:3.12}
ds^2 = - f(r)dt^2_r + 2 \sqrt{\frac{2GM}{r} + \frac{2 \tilde{\beta} G^2 M}{r(r^2 - R^2_s)}} dt_r dr + dr^2 +r^2 d\Omega^2
\end{equation}
In the following part of section 3, we will take $dt^2 \equiv dt^2_r$ for convenience, i.e.,
\begin{equation}\label{eq:3.13}
ds^2 = - f(r)dt^2 + 2 \sqrt{\frac{2GM}{r} + \frac{2 \tilde{\beta} G^2 M}{r(r^2 - R^2_s)}} dt dr + dr^2 +r^2 d\Omega^2
\end{equation}
where $f(r)$ has been given in eq.$(2.5)$.\\
Furthermore, we can expand eq.$(3.11)$ in powers of $G$. In order to compare with the Schwarzschild metric easily, then the metric $(3.13)$ is approximate to
\begin{equation}\label{eq:3.14}
ds^2 = - f(r)dt^2 + 2 \sqrt{\frac{2GM}{r}} \left(1 + \frac{\tilde{\beta} G}{2(r^2 - R^2_s)}\right) dt dr + dr^2 +r^2 d\Omega^2
\end{equation}
The action for a particle moving freely in a curved background can be written as
\begin{equation}\label{eq:3.15}
S = \int p_{\mu} dx^{\mu}
\end{equation}
with 
\begin{equation}\label{eq:3.16}
p_{\mu} = m g_{\mu\nu} \frac{dx^{\nu}}{d\sigma}
\end{equation}
where $\sigma$ is an affine parameter along the worldline of the particle, chosen so that $p_{\mu}$ coincides with the physical 4-momentum of the particle. For a massive particle, this requires that $d\sigma = d\tau/m$, with $\tau$ the proper time [9].\\
For simplicity, we only discuss the excitations for a massless scalar field $\Phi$ in this article and ignore angular directions in section 3. Following the spirit of ref.[9], we will present a calculation of the imaginary part of the action for a massless point particle tunnelling through the horizon of Schwarzschild spacetime in section 3. \\

\subsection{Quantum correction to Hawking temperature}
The radial dynamics of massless particles in Schwarzschild spacetime are determined by the equations
\begin{equation}\label{eq:3.17}
f(r) \dot{t}^2 - 2 \sqrt{\frac{2GM}{r}} \left(1 + \frac{\tilde{\beta} G}{2(r^2 - R^2_s)}\right) \dot{t} \dot{r} - \dot{r}^2  = 0 
\end{equation}
\begin{equation}\label{eq:3.18}
f(r) \dot{t} -  \sqrt{\frac{2GM}{r}} \left(1 + \frac{\tilde{\beta} G}{2(r^2 - R^2_s)}\right)  \dot{r}  = \omega 
\end{equation}
The second equation is the geodesic equation corresponding to the time-independence of the metric; in terms of the momentum defined in eq. $(3.16)$, it can be written $p_t = -\omega$, and so $\omega$ has the interpretation of the energy of the particle as measured at infinity [9].\\
We now consider the the trajectory of an outgoing particle. Classical outgoing trajectory which crosses the horizon is forbidden. However, the analytic continuation of an outgoing trajectory with $r > 2M$ backwards across the horizon will give rise to an imaginary term in our action, and this term represents the tunnelling amplitude [9]. Eq.$(3.17)$ can be factorised to the equation:
\begin{equation}\label{eq:3.19}
\frac{dr}{dt} \approx \pm (1- \sqrt{\frac{2GM}{r}}) - \sqrt{\frac{2GM}{r}} \frac{\tilde{\beta} G}{2(r^2 - R^2_s)} 
\end{equation}
with the upper (lower) sign in eq.$(3.19)$ corresponding to outgoing (ingoing) geodesics, under the implicit assumption that $t$ increases towards the future [6]. As we have pointed out, in section 3, we only consider the case of upper sign, i.e., outgoing particles.
Considering eq.$(3.19)$, we then have
\begin{equation}\label{eq:3.20}
Im S = Im \int p_r dr = Im \int [ \sqrt{\frac{2GM}{r}} \left(1 + \frac{\tilde{\beta} G}{2(r^2 - R^2_s)}\right) \dot{t} + \dot{r}] dr = Im \int \dot{t} dr
\end{equation}
Combining eqs.$(3.18)$ and $(3.19)$, we have
\begin{equation}\label{eq:3.21}
\dot{t} = \frac{\omega}{1 - \sqrt{\frac{2GM}{r}} - \sqrt{\frac{2GM}{r}} \frac{\tilde{\beta} G}{2(r^2 - R^2_s)} + \frac{1}{4} K(r) G^2}
\end{equation}
where $K(r) = \frac{4}{G^2} \left(\alpha_e(r) + \beta_e (r) \right)$. To calculate eq.$(3.20)$, we can define $\tilde{M}$ which satisfies the following relation
\begin{equation}\label{eq:3.22}
1- \sqrt{\frac{2G \tilde{M}}{r}} = 1- \sqrt{\frac{2G M}{r}} - \sqrt{\frac{2G M}{r}} \frac{\tilde{\beta} G}{2(r^2 - R^2_s)} + \frac{1}{4} K(r) G^2 
\end{equation}
namely,
\begin{equation}\label{eq:3.23}
\sqrt{\frac{2G \tilde{M}}{r}} = \sqrt{\frac{2G M}{r}} + \sqrt{\frac{2G M}{r}} \frac{\tilde{\beta} G}{2(r^2 - R^2_s)} - \frac{1}{4} K(r) G^2
\end{equation}
Then eq.$(3.20)$ can be written as
\begin{equation}\label{eq:3.24}
Im S = Im \int \dot{t} dr = \int \frac{\omega}{1- \sqrt{\frac{2G \tilde{M}}{r}}}
\end{equation}
The integrand has a pole at $r_h = 2G\tilde{M}$, the horizon. Choosing the prescription to integrate clockwise around this pole (into the upper-half complex-$r$ plane) [9], we find
\begin{equation}\label{eq:3.25}
Im S = 4\pi G \tilde{M} \omega 
\end{equation}
Then according eq.$(3.3)$, we have the temperature at which the black hole radiates
\begin{equation}\label{eq:3.26}
T_{BH} = \frac{1}{8 \pi G \tilde{M}}
\end{equation}
From $(3.23)$, we can obtain the expression of $\tilde{M}$: 
\begin{equation}\label{eq:3.27}
\tilde{M} = M + \frac{\tilde{\beta} G}{r^2 - R^2_s} M + \frac{\tilde{\beta}^2 G^2}{4(r^2 - R^2_s)} M - \frac{1}{4} K(r) r \sqrt{\frac{2GM}{r}}G
\end{equation}
Furthermore, if the last term of $(3.27)$ can be neglected, which is the case we mainly consider, then we have
\begin{equation}\label{eq:3.28}
\tilde{M} = M + \frac{\tilde{\beta} G}{r^2 - R^2_s} M + \frac{\tilde{\beta}^2 G^2}{4(r^2 - R^2_s)} M = \left(1+ \frac{\tilde{\beta} G}{2(r^2-R^2_s)} \right)^2 M
\end{equation}
Considering eqs.$(3.25)$ and $(3.26)$, the imaginary part of the action $ImS$ and the Hawking temperature $T_{BH}$ are given by
\begin{equation}\label{eq:3.29}
ImS \approx 4 \pi G M \omega + \frac{4\pi \tilde{\beta} G^2 M \omega}{r^2 - R^2_s} + O(G^3)
\end{equation}
\begin{equation}\label{eq:3.30}
T_{BH} \approx \frac{1}{8 \pi G M}\left( 1- \frac{\tilde{\beta} G}{r^2 - R^2_s} - \frac{\tilde{\beta}^2 G^2}{4(r^2 - R^2_s)} \right)
\end{equation}
These results can be compared with that in section 4.\\

\subsection{Quantum correction to Hawking entropy}
On the one hand, according to thermodynamics, we have $dS_{BH} = \frac{dM}{T_{BH}} = 8 \pi G \tilde{M} dM$, where $S_{BH}$ represents the Hawking entropy of black holes. We notice that $M = \frac{4\pi}{3} R^3_s \rho_0$, where $\rho_0$ is the average density of black holes [12,21]. On the one hand, if we fix the radius $R_s$ and allow $\rho_0$ to vary, then we can obtain the quantum corrected Hawking entropy to order $G^2$
\begin{equation}\label{eq:3.31}
S_{BH} =4 \pi G M^2 + \frac{4\pi \tilde{\beta} G^2 M \omega}{r^2 - R^2_s} 
\end{equation}
On the other hand, if we fix the radius $\rho_0$ and allow $R_s$ to vary, then to order $G^2$, the quantum corrected Hawking entropy should be
\begin{equation}\label{eq:3.32}
S_{BH} =4 \pi G M^2 + \frac{64}{3}\pi^3 \tilde{\beta} G^2 \rho^2_0 [-r^4 \ln(\frac{r^2 - R^2_s}{r^2}) + \frac{1}{2} r^4 - \frac{1}{2} (r^2 - R^2_s)^2 - 2R^2_s r^2] 
\end{equation}
Moreover, we can also obtain the area of horizon at $ r_h = 2G \tilde{M}$
\begin{equation}\label{eq:3.33}
A = 4 \pi r^2_h = 16 \pi^2 G^2 M^2 + O(G^3)
\end{equation}
Comparing eqs.$(3.28)$, $(3.29)$ and $(3.30)$, we can find that the classical area law of black holes $S= \frac{1}{4G} A$ should be modified.\\
Furthermore, the heat capacity of black holes is given by
\begin{equation}\label{eq:3.34}
C = T_{BH} \frac{\partial S}{\partial T_{BH}} = -8\pi G (1 + \frac{\tilde{\beta} G}{2(r^2 - R^2_s)})^2 M^2 < 0
\end{equation}
Therefore, thermal equilibrium cannot always exist between black holes and external environment.\\
We can find that the results in eqs.$(3.28)-(3.34)$ all depend on $R_s$. However, in order to ensure that the quantum modifications are tiny, then the term $\frac{\tilde{\beta} G}{2(r^2 - R^2_s)}$ in $(3.28)$ should be very small, namely, $r \gg R_s$. If $R_s$ can be neglected, then the results in eqs.$(3.28)-(3.34)$ do not depend on $R_s$. However, it is easy to check that the main conclusions will not change, namely, the classical area law of black holes $S= \frac{1}{4G}A$ should be modified and the heat capacity of black holes is always negative.

\section{Hawking radiation as tunneling for Schwarzschild black holes}
In this section, we will think of Hawking radiation as tunneling along another method. Maulik K. Parikh and Frank Wilczek [6] have written a famous paper on this topic. The authors found that the exact spectrum of Hawking radiation is not precisely thermal [6]. Therefore, information can be conserved. Following the spirit of ref.[6], We present a short and direct derivation of Hawking radiation as a tunneling
process, based on particles in the quantum corrected metric $(2.4)$. \\
The radial null geodesics have been given by eq.$(3.19)$, namely,
\begin{equation}\label{eq:4.1}
\frac{dr}{dt} \approx \pm (1- \sqrt{\frac{2GM}{r}}) - \sqrt{\frac{2GM}{r}} \frac{\tilde{\beta} G}{2(r^2 - R^2_s)} 
\end{equation}
However, these equations should be modified if we take into account the effect of the particle’s self-gravitation. Self-gravitating shells in Hamiltonian gravity were studied by Kraus and Wilczek [22]. The authors found that when the black hole mass is held fixed and the total ADM mass allowed to vary, a shell of energy $\omega$ moves in the geodesics of a spacetime with $M$ replaced by $M + \omega$ [6,22]. Therefore, if we fix the total mass and allow the black hole mass to vary, then the shell of energy $\omega$ travels on the geodesics given by the line element
\begin{equation}\label{eq:4.2}
ds^2 = -f(r, M-\omega) dt^2 + 2 \sqrt{\frac{2G(M-\omega)}{r} + \beta_e(r, M-\omega)} dt dr + dr^2,
\end{equation}
where 
\begin{equation}\label{eq:4.3}
f(r, M-\omega) = 1 - \frac{2G(M-\omega)}{r} + \alpha_e(r,M-\omega)
\end{equation}
\begin{equation}\label{eq:4.4}
\alpha_e(r, M-\omega)= \tilde{\alpha} \frac{2G^2(M-\omega)}{R^3_s} [2\frac{R_s}{r} + \ln (\frac{r-R_s}{r+R_s})] + O(G^3)
\end{equation}
\begin{equation}\label{eq:4.5}
\beta_e(r, M-\omega)= \tilde{\beta} \frac{2G^2(M-\omega)}{r (r^2 - R^2_s)} + O(G^3)
\end{equation}
and we have ignored the angular directions. So we should use eq.$(4.1)$ with $M \rightarrow M-\omega$.\\
The imaginary part of the action for an $s$-wave outgoing positive energy particle which crosses the horizon outwards from $r_{in}$ to $r_{out}$ can be expressed as [6]
\begin{equation}\label{eq:4.6}
ImS= Im \int^{r_{out}}_{r_{in}} p_r dr = Im \int^{r_{out}}_{r_{in}} \int^{p_r}_0 dp'_r dr
\end{equation}
Then we use Hamilton's equation $\dot{r} = + \frac{dH}{dp_r} \vert_{r}$, change variable from momentum to energy, and switch the order of integration to obtain [6]
\begin{equation}\label{eq:4.7}
\begin{aligned}
ImS= Im \int^{M-\omega}_{M} \int^{r_{out}}_{r_{in}}  \frac{dr}{\dot{r}} dH = 
-Im \int^{\omega}_0 d\omega' \int^{r_{out}}_{r_{in}} \frac{dr}{\left(1- \sqrt{\frac{2G(M-\omega')}{r}} \right) - \sqrt{\frac{2G(M-\omega')}{r}} \frac{\tilde{\beta} G}{2(r^2 - R^2_s)}}
\end{aligned}
\end{equation}
where we have used the modified eq.$(4.1)$, and the minus sign appears from $H = M-\omega'$.\\
We can define the same $\tilde{M}$ as in section 3.1, namely,
\begin{equation}\label{eq:4.8}
ImS= Im \int^{M-\omega}_{M} \int^{r_{out}}_{r_{in}}  \frac{dr}{\dot{r}} dH = 
-Im \int^{\omega}_0 d\omega' \int^{r_{out}}_{r_{in}} \frac{dr}{1- \sqrt{\frac{2G (\tilde{M} - \omega')}{r}}}
\end{equation} 
The integral can be done by deforming the contour, so as to ensure that positive energy solutions decay in time, i.e., into the lower half $\omega'$ plane [6]. Then we obtain 
\begin{equation}\label{eq:4.9}
ImS= 4 \pi G M \omega + \frac{4\pi \tilde{\beta} G^2 M \omega}{r^2 - R^2_s} - 2\pi G \omega^2 + O(G^3)
\end{equation} 
It is obvious that if $\omega^2$ is very small, the result eq.$(3.29)$ would be recovered. In other words, the result which we have obtained in section 3.1 is reasonable.\\
Here we should know that the derivation in section 3 and section 4 are slightly different. In section 3, we have neglected the back-reaction of the radiation on the black hole geometry, since we are only interested in the nature of the spectrum to leading
order [9]. However, in section 4 the particle’s self-gravitation has been taken into account [22]. The authors found that when the black hole mass is held fixed and the total ADM mass allowed to vary, a shell of energy $\omega$ moves in the geodesics of a spacetime with $M$ replaced by $M + \omega$ [6,22]. Therefore, the mass $M$ in eqs. $(4.1) - (4.5)$ should be replaced by $M-\omega$. These are the main differences of derivation between section 3 and section 4. \\
For classical Schwarzschild black holes, Maulik K. Parikh and Frank Wilczek [6] have found that  
\begin{equation}\label{eq:4.10}
ImS= 4\pi G M \omega (1 - \frac{\omega}{2M})
\end{equation}
Only when the energy of emitted particles $\omega$ is very low, namely, $\omega^2$ can be neglected, pure blackbody spectrum can be recovered. If the energy of emitted particles $\omega$ is very high, blackbody spectrum would be deviated. This result satisfies unitarity and supports information conservation.\\
However, when
\begin{equation}\label{eq:4.11}
\omega = \frac{2\tilde{\beta} G M}{r^2 - R^2_s},
\end{equation}
eq.$(4.9)$ will become pure blackbody spectrum. In particular, for $r \gg R_s$, $\omega = 2\tilde{\beta} G M/r^2$. Therefore, for the quantum corrected metric, it is possible to break down information conservation. Therefore, we suggest that, in order to ensure conservation of information at any energy scales, black hole needs to retain some “remnant”, namely, a black hole cannot release all information it has as Hawking suggested [26].\\
Of course, all the above discussions are only true for quasi-static reversible processes. However, since the heat capacity of black holes is negative (i.e., eq.$(3.31)$), thermal equilibruim cannot always exist between black holes and external environment. For irreversible processes, further studies are needed.\\  
Careful readers can find that since $ImS$ can be obtained by integrating along the trajectory, such as eqs.$(3.20)$ and $(4.7)$, the results should not depend on the generic radial coordinate $r$. However, the results in eqs.$(3.28)-(3.30)$ and $(4.9)$ all depend on radial coordinate $r$. So, what are the differences between these two arguements? The explanation consists of three parts.\\
Firstly, the integrations in eqs.$(3.20)$ and $(4.7)$ are hard to obtain the exact expressions (although can be obtained by Mathematica). So, in this article, we do not hope to obtain the exact formulae. However, in section 3 and 4 we have used slightly different methods respectively to obtain the same results for $ImS$ when $\omega^2$ can be neglected. Therefore, we can guarantee that the methods and results in section 3 and 4 are reasonable and consistent.\\
Secondly, even if we cannot obtain the exact formulae, the main results in section 3 and 4 are remained when quantum correction has been considered. The first one is that the famous area law of black holes $S=\frac{1}{4G}A$ should be corrected. The second one is that the heat capacity of black holes is negative, which means that thermal equilibrium cannot always exist between black holes and external environment. The third one is that at certain frequency $\omega$, the radiation spectrum will become pure blackbody spectrum and in order to ensure information conservation a black hole needs to retain some “remnant”. \\
Thirdly, the results in eqs.$(3.28)-(3.30)$ and $(4.9)$ can be easily compared with the results for Schwarzschild solution.  

\section{Bound on greybody factors for Schwarzschild black holes}
In section 3 and 4, we have found that Hawking radiation spectrum of black holes deviates from blackbody spectrum in most cases for the quantum corrected metric. Therefore, it is natural to discuss greybody factors $\sigma_l(\omega)$ for Schwarzschild black holes. The total radiation power is given by
\begin{equation}\label{eq:5.1}
P(\omega) = \sum_l \int^{\infty}_0 P_l (\omega) d\omega,
\end{equation} 
where  
\begin{equation}\label{eq:5.2}
P_l (\omega) = \frac{A}{8\pi^2} \sigma_l (\omega) \frac{\omega^3}{\exp(\omega/T_{BH}) -1}
\end{equation} 
is the power emitted per unit frequency in the $l^{th}$ mode, $A$ is a a multiple of the horizon area, $l$ is the angular momentum quantum number and $\sigma_l (\omega)$ is the frequency dependent greybody factor [10]. One finds that if the black hole is to be in thermal equilibrium with a bath of particles surrounding it, the rate at which it emits these particles must equal the rate at which particles in the bath fall into the black hole [9]. \\
There are various methods in literature [27–31] to estimate the greybody factors, however, derivation of exact analytical expression of greybody factor is limited only to very few cases [10]. Following ref.[32], we will briefly consider the bounds on the greybody factor for the quantum corrected black hole metric. The general bounds on the greybody factor, as proposed by Petarpa Boonserm and Matt Visser [32] is given by
\begin{equation}\label{eq:5.3}
\sigma_l (\omega) \geq sech^2 \left(\int^{\infty}_{-\infty} \Theta dr_{\ast} \right)
\end{equation} 
where
\begin{equation}\label{eq:5.4}
\Theta = \frac{\sqrt{[h'(r)^2] + [\omega^2- V_{eff} - h(r)^2]^2 }}{2h(r)}
\end{equation}
The arbitrary function $h(r)$ has to be positive definite everywhere and satisfy the boundary condition, $h(\infty) =h(-\infty) = \omega$ for the bound eq.$(5.2)$ to hold [3] and $V_{eff}$ is the effective potential which will be given later. \\
The equation of motion for a free massless scalar field $\Phi$ is given by
\begin{equation}\label{eq:5.5}
\Box \Phi = 0
\end{equation}
Since metric $(2.12)$ has spherical symmetry, the angular variables can be seperated from the other coordinates, i.e.[21],
\begin{equation}\label{eq:5.6}
\Phi (t,r,\theta,\phi) = \Phi(t,r) S(\theta,\phi) 
\end{equation}
where $S(\theta,\phi)$ can be decomposed in the usual spherical harmonics satisfying [21]
\begin{equation}\label{eq:5.7}
\left(\partial^2_{\theta} + \frac{\cos \theta}{\sin \theta} \partial_{\theta} + \frac{1}{\sin^2 \theta} \partial^2_{\phi} \right) Y(\theta,\phi) = -l(l+1) Y(\theta,\phi)  
\end{equation}
Moreover, time $t$ can be seperated from the radial coordinate, i.e.,
\begin{equation}\label{eq:5.8}
\Phi(t,r) = \Psi (t) \Phi(r) 
\end{equation}
where $\Psi (t) \sim e^{i \omega t}$ and satisfies $\ddot{\Psi} (t) = - \omega^2 \Psi(t)$ [21].\\
Furthermore, cosidering the metric $(2.12)$, the radial equation for a massless scalar field $\Phi$ with the tortoise like coordinate $r_{\ast}$, which has been given by eq.$(2.11)$, is [21]
\begin{equation}\label{eq:5.9}
[\partial^2_{r_{\ast}} + \omega^2]u(r) = V_{eff} u(r) 
\end{equation}
where 
\begin{equation}\label{eq:5.10}
V_{eff} = f(r) [\frac{l(l+1)}{r^2} + \frac{2GM}{r^3} ]
\end{equation}
and $l$ is the angular momentum of the specific wave mode under consideration [10].\\
A particularly simple choice of $h(r)$ for the present case is [3,10,32]
\begin{equation}\label{eq:5.11}
h(r) = \omega
\end{equation}
Combining eqs.$(2.11)$, $(2.13)$, $(5.3)$ and $(5.4)$, if we only consider the case of lowest order, then 
\begin{equation}\label{eq:5.12}
\sigma_l (\omega) \geq sech^2 \left(\frac{1}{2\omega} \int^{\infty}_{r_{+}} [\frac{l(l+1)}{r^2} + \frac{2GM}{r^3}] dr \right)
\end{equation}
where $r_{+}$ is the modified horizon at $r= 2G \tilde{M}$. Therefore,
\begin{equation}\label{eq:5.13}
\sigma_l (\omega) \geq sech^2 \left(\frac{l(l+1)}{4G \tilde{M} \omega} + \frac{M}{8G \tilde{M}^2 \omega} \right)
\end{equation}
For classical Schwarzschild black holes, the horizon is at $r=2GM$, then the bound of greybody factors is [32]
\begin{equation}\label{eq:5.14}
\sigma_l (\omega) \geq sech^2 \left(\frac{1}{2\omega} \int^{\infty}_{2GM} [\frac{l(l+1)}{r^2} + \frac{2GM}{r^3}] dr \right)
\end{equation}
namely,
\begin{equation}\label{eq:5.15}
\sigma_l (\omega) \geq sech^2 \left(\frac{l(l+1)}{4G M \omega} + \frac{1}{8G M \omega} \right)
\end{equation}
Considering eq.$(3.24)$ and comparing eqs.$(5.13)$ and $(5.15)$, it is easy to find out the bound on greybody factors has a very small difference between classical metric and quantum corrected metric. It can be expected that the modification of total radiation power will be very small as well.

\section{Conclusions and Discussions}
Since 1974, Hawking radiation [1,2] has become a fascinating research field since it relates classical general relativity and quantum field theory and is a deep insight into quantum gravity. Many researchers [4-8] devoted to calculating the Hawking temperature and radiation power for exploring a possible quantum gravity theory [3]. Furthermore, these methods have been generalized to other cases, such as charged black holes [3,10], Einstein-Gauss-Bonnet de Sitter black hole [11], etc. However, the quantum effect on metrics have been neglected in all mentioned studies.\\
Many researches have been conducted in quantum gravity using effective field theory (EFT) [13-20]. Though finding a consistent quantum gravity theory is still an extremely difficult challenge at the moment, one can still apply EFT methods to do calculations in quantum gravity which are model independent at energies below the Planck mass [12,21]. Xavier Calmet, Roberto Casadio and Folkert Kuipers [12] have used EFT methods [12,21] to obtain the metric inside and outside a star, taking into account quantum effects. Moreover, in ref.[12] the authors also proposed that the exterior metric may provide a good model of quantum black holes. Therefore, one direct research step to consider the effect of quantum correction on Hawking radiation. For simplicity, we only discuss the excitations for a massless scalar field $\Phi$ in this article.\\
In section 3, we have studied quantum correction of Hawking temperature and entropy for Schwarzschild black holes. Firstly, we studied the radial dynamic equations of massless particles in quantum corrected Schwarzschild spacetime. Then, following the spirit of ref.[9], we obtained the modified Hawking temperature. Moreover, by calculating the modifed Hawking entropy and the area of horizon, we found that the famous area law of black holes $S=\frac{1}{4G}A$ need to be corrected. Some researches have been conducted to study the possibility to generalize area law of black holes to more general forms [33,34]. For example, if we consider non-extensive statistical mechanics, the area law should be written as $S= \gamma A^{\beta}$, where $A$ is the black hole horizon area, $\gamma$ is an unknown constant and $\beta$ known as non-extensive parameter which is a constant [34]. However, for the case we have studied in this article, $\gamma=\frac{1}{4G}$ is a constant while the parameter $\beta$ is dependent on $A$ or $M$. More future studies should focus on this difference and relationship. Furthermore, we have determined the heat capacity of Schwarzschild black holes as well. Even if the quantum corrections have been considered, the heat capacity of black holes is negative, which means that thermal equilibruim cannot always exist between black holes and external environment.\\
In section 4, we considered Hawking radiation as tunneling following the methods in ref.[6]. To leading order, we obtained the same results in section 3.1, therefore, the results are reasonable. For most cases, when the energy of emitted particles $\omega$ is very low, pure blackbody spectrum can be recovered. While if the energy of emitted particles $\omega$ is very high, blackbody spectrum would be deviated. This result is as the same as the result of ref.[6] for classical Schwarzschild black holes. However, there exits a special case for the quantum corrected metric. When the frequency of emitted particles takes certain value (in this article $\omega = \frac{2\tilde{\beta} G M}{r^2 - R^2_s}$), the radiation spectrum will become pure blackbody spectrum. Therefore, it has the possibility to break down information conservation. In order to ensure the conservation of information at any energy scales, we suggested that a black hole needs to keep some information it has. We should point out that all the results in section 4 are only true for quasi-static reversible processes. Since the heat capacity of black holes is negative, thermal equilibrium cannot always exist between black holes and external environment. Further studies for irreversible processes are needed.\\
In section 5, we briefly discussed the bound on greybody factors for Schwarzschild black holes to lowest order and found that the bound has a very small difference between classical metric and quantum corrected metric.\\
Future work can be directed along at least three lines of further research. Firstly, the quantum corrected metric for other classical solutions of black holes should be obtained using EFT methods. Secondly, we can generalize the methods and results in this article to other quantum corrected black hole metrics. Thirdly, we can apply the methods in refs.[6,9] to study the properties of Hawking radiation in stringy gravity and test if information is conserved or not. There is therefore great potential for development of this work in the future.



\end{document}